\documentclass[prb,twocolumn,showpacs,preprintnumbers,amsmath,amssymb]{revtex4-1}
\usepackage{amssymb}
\usepackage{bm}
\usepackage{amsmath}
\usepackage[dvips]{graphicx}
\usepackage{color}
\usepackage{comment}
\usepackage{multirow}
\usepackage{subfigure}

\begin{document}
\color{red}
\title{Two-dimensional topological semimetal state in a nanopatterned semiconductor system}
\color{black}

\author{Tommy Li$^1$ and Oleg P. Sushkov$^1$}
\affiliation{$^1$School of Physics, University of New South Wales, Sydney 2052, Australia}
\begin{abstract}
We propose the creation of a two-dimensional topological semimetal in a semiconductor artificial lattice with triangular symmetry. An in-plane magnetic field drives a quantum phase transition  between the topological insulating and topological semimetal phases. The topological semimetal  is characterized by robust band touching points which carry quantized Berry flux and edge states which terminate at the band touching points. The topological phase transition is predicted to occur at magnetic fields $\sim 4\text{T}$ in high mobility GaAs artificial lattices, and can be detected via the anomalous behaviour of the edge conductance.
\end{abstract}

\maketitle

\begin{figure*}
\begin{tabular}{ccc}
\includegraphics[width = 0.3\textwidth]{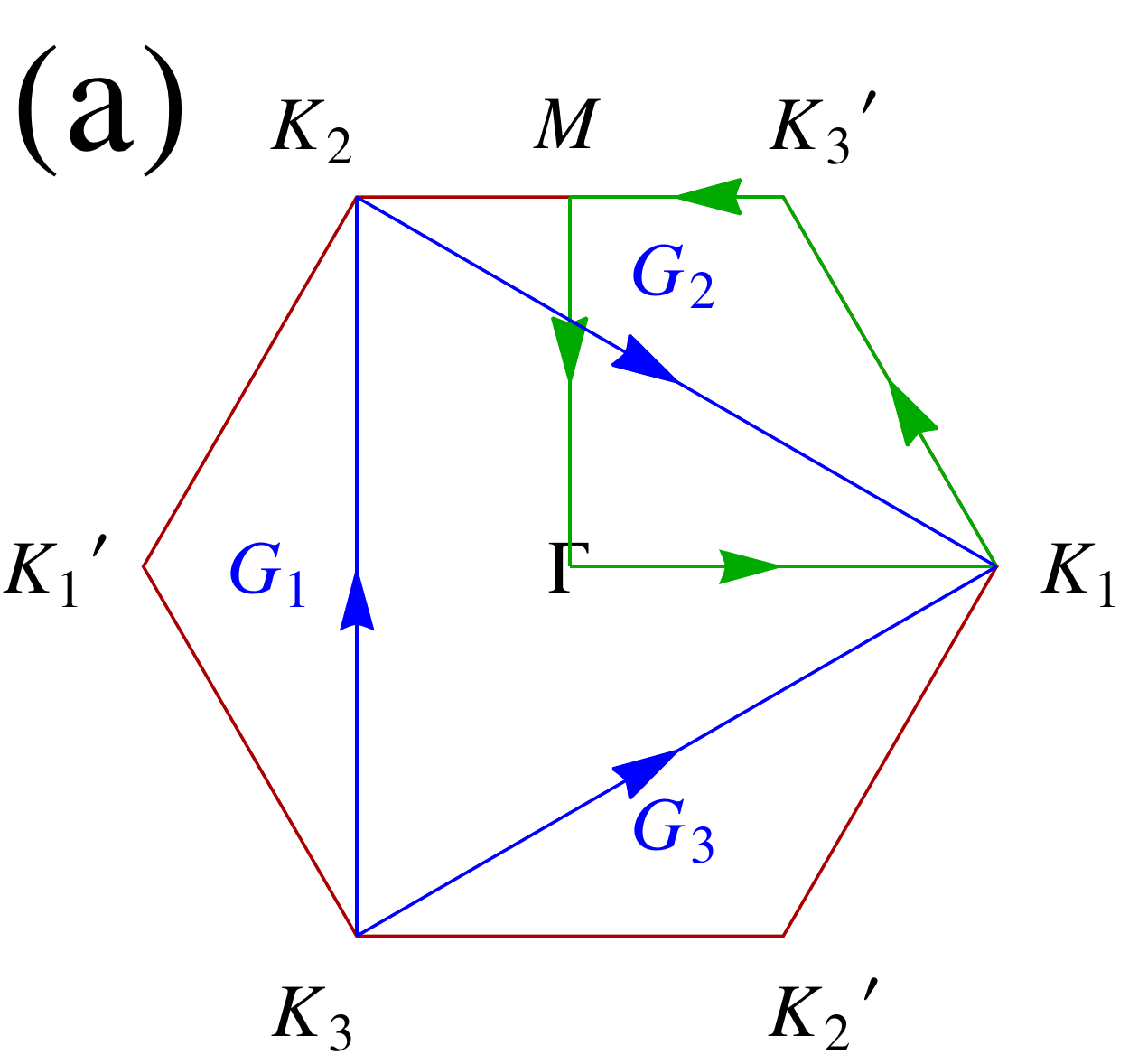}
& ~ ~ &
\includegraphics[width = 0.4\textwidth]{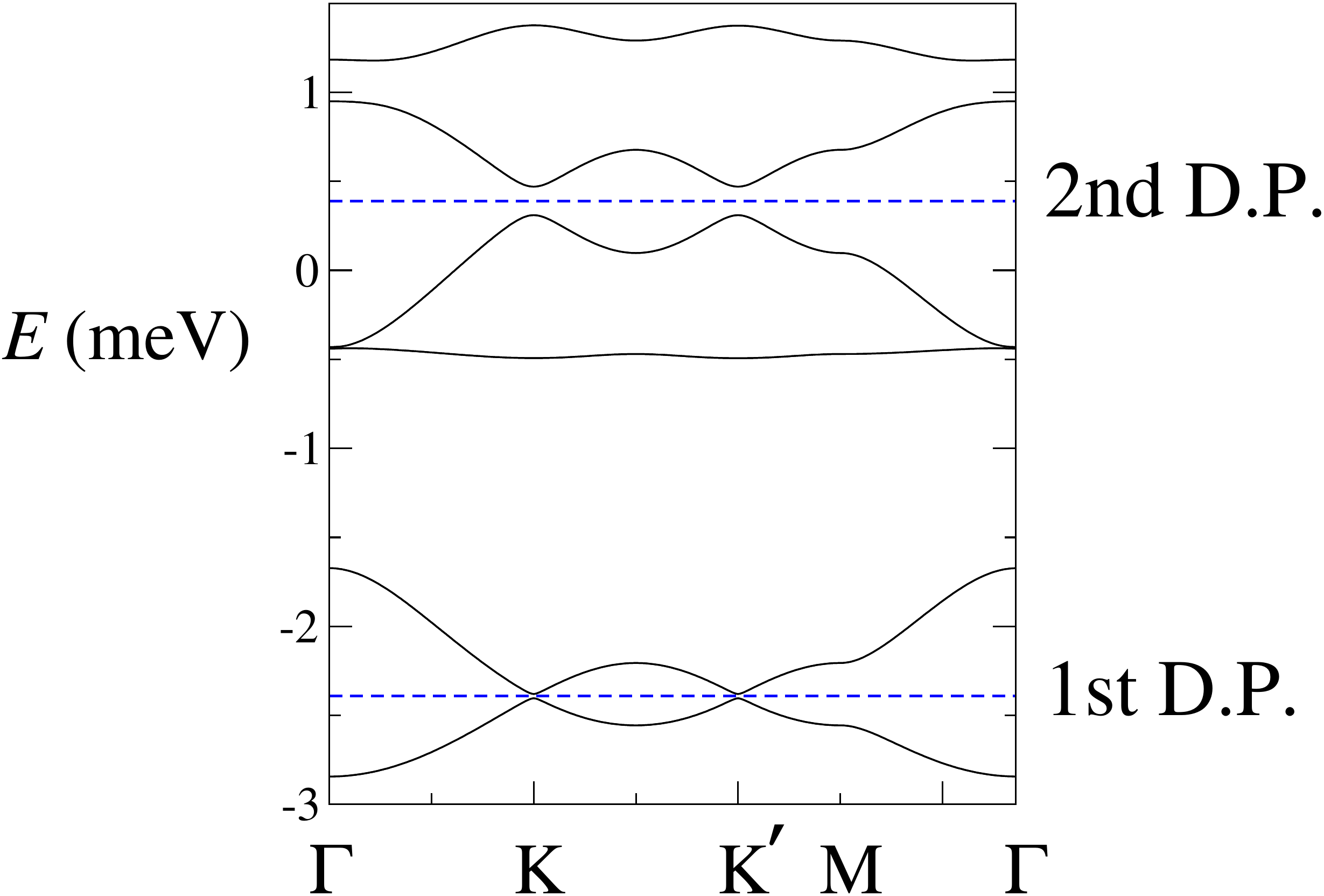}
\end{tabular}
\caption{The Brillouin zone (left) and typical dispersion (right) of a hexagonally patterned semiconductor hole gas at zero magnetic field with superlattice potential $U(x, y)$ given by Eq. (\ref{latt}). The dispersion in panel b is shown along the path indicated by the green arrows in panel a for lattice spacing $a = 60\text{nm}$, well width $d = 20\text{nm}$ and $W = 2 E_0$. The dashed lines in panel b indicate the chemical potential at the 1st and 2nd Dirac points, at which the system is in a topological insulating state.}
\end{figure*}

The discovery of two-dimensional (2D) and three-dimensional (3D) topological insulators\cite{KaneMele,QSHobs,MooreBalents,TI3Dobs} has spurred on efforts to classify insulating states of arbitrary dimension according to their bulk topological invariants\cite{Schnyder,Kitaev}. Topologically non-trivial insulating states simultaneously possess quantized bulk transport coefficients and protected edge modes\cite{topquant,bulkboundary}. In two and three dimensions, these insulating phases are analogues of the Quantum Hall system\cite{TKNN}, of which numerous examples have been experimentally realized or observed  including Quantum Spin Hall insulators\cite{QSHobs,QSHobsInAs}, Quantum Anomalous Hall Insulators\cite{QAHobs} and 3D topological insulators\cite{TI3Dobs}. Since the non-trivial topology of these systems relies on the existence of a gap in the bulk dispersion, it is remarkable that a class of 3D topological gapless systems has been both theoretically proposed and experimentally observed\cite{Weyltheory,Weylobs}. These systems realize Weyl fermion  physics\cite{Weylfermion}, possessing degeneracies in the bulk spectrum which cannot be gapped by arbitrary perturbations to the band structure. Weyl semimetals are phenomenologically characterized by the presence of a chiral anomaly in magnetotransport\cite{WeylChiraltheory,WeylChiralobs} and topologically protected \emph{Fermi arc} surface states which connect the Weyl points in momentum space\cite{Weyltheory,Weylobs}. More recently, 2D semimetals with symmetry-protected band touching points have been considered\cite{TMDC}, and a  2D topological semimetallic phase has been proposed in an optical lattice\cite{OptLatt} . In this work we propose the realization of the 2D topological semimetal in a hexagonally patterned semiconductor system. The band touching points in our system carry quantized Berry flux and cannot be gapped by arbitrary spin-orbit interactions. The quantum phase transition between the topological insulator \cite{SushkovCastroNeto} and  topological semimetal states is driven by an in-plane magnetic field. At the phase transition the one-dimensional (1D) edge states behave analogously to the Fermi arc states in 3D Weyl semimetals, terminating at the band touching points. This behaviour provides an experimental signature of the topological phase transition via  measurement of the  longitudinal conductance.

The system we propose is a hexagonally patterned 2D hole gas in a $p$-type zincblende semiconductor quantum well, which is described by the 3D Hamiltonian\cite{Luttinger}
\begin{gather}
H_{3D} = \frac{1}{2m_e} ( (\gamma_1 + \frac{5}{2}\gamma_2) p^2 - \gamma_2 (\bm{p} \cdot \bm{S})^2) + H' \nonumber \\
+ V(z) + U(x,y) - 2\kappa \mu_B (\bm{B} \cdot \bm{S}) \ \ ,
\label{hamil3D}
\end{gather}
which is constructed from spin-$\frac{3}{2}$ operators $\bm{S}$. The first term in (\ref{hamil3D}) is the isotropic contribution to the bulk Hamiltonian of the crystal, with $\gamma_1, \gamma_2$ being the Luttinger parameters\cite{Luttinger}, and $H'$ includes all remaining terms in the bulk Hamiltonian, including crystallographic anisotropy and bulk inversion asymmetric interactions. $V(z)$ and $U(x,y)$ represent the 2D confining potential and the hexagonal lattice potential respectively, and the final term describes the Zeeman coupling to a magnetic field $\bm{B}$ applied in the 2D plane. Hereafter we assume that confinement occurs along a high-symmetry direction of the crystal, and take coordinates in which $x, y$ are directed in the 2D plane\cite{footnote}. The presence of the magnetic field violates time reversal symmetry, however we may construct an exact antiunitary symmetry of the Hamiltonian (\ref{hamil3D}). Let us introduce an antiunitary operator $\widetilde{\mathcal{T}} = \Omega \mathcal{K}$ (with $\mathcal{K}$ being complex conjugation), where $\Omega$ is a tensor operator in spin space satisfying $\Omega| \pm \frac{3}{2} \rangle = |\mp \frac{1}{2}\rangle$, $\Omega|\pm \frac{1}{2} \rangle = |\mp \frac{3}{2}\rangle$. Under the action of $\widetilde{\mathcal{T}}$ the spin and momentum operators transform as
\begin{gather}
S_x \rightarrow S_x \ \ , \ \ S_y \rightarrow S_y \ \ , \ \ S_z \rightarrow - S_z \ \ , \ \ \bm{p} \rightarrow - \bm{p} \ \ .
\end{gather}
Note that $\widetilde{\mathcal{T}}^2 = +1$. Due to the tetrahedral symmetry of zincblende crystals, the complete band Hamiltonian consists of terms which either contain odd powers of both $k_z$ and $S_z$ or even powers of both operators. If the in-plane potential satisfies $U(x, y) = U(-x, -y)$, it follows that the Hamiltonian (\ref{hamil3D}) is invariant under the symmetry operation $P\widetilde{\mathcal{T}}$ where $P$ is inversion in the plane, $x \rightarrow -x \ \ , \ \ y \rightarrow - y  \ \ , \ \ z \rightarrow +z$, even in the presence of spin-orbit interactions associated with bulk inversion asymmetry (e.g. the Dresselhaus interaction) or asymmetry of the confining potential $V(z)$ (e.g. the Rashba interaction). 

We will consider the explicit in-plane potential
\begin{gather}
U(x, y) = 2W ( \cos \bm{G}_1 \cdot \bm{r} + \cos \bm{G}_2 \cdot \bm{r} + \cos \bm{G}_3 \cdot \bm{r})
\label{latt}
\end{gather}
which describes an artificial hexagonal lattice, with $\bm{G}_1, \bm{G}_2, \bm{G}_3$ being the reciprocal lattice vectors shown in Fig. 1a. The lattice potential satisfies inversion symmetry in the plane, $U(x, y) = U(-x, -y)$. We assume $W > 0$ so the potential has maxima at the lattice sites. The bandwidth is approximately given by $E_0 = \frac{K^2}{2m}$ where $K = \frac{ 4\pi}{3}$ is the momentum at the corners of the Brillouin zone and $m$ is the effective mass of 2D holes. The dispersion  for parameters $a = 60\text{nm}, d = 20\text{nm}, W = 2E_0$ is shown in Fig. 1b. At zero magnetic field the system simulates the Kane-Mele model\cite{KaneMele} of graphene with spin-orbit interaction.\cite{SushkovCastroNeto}. The system is in a topological insulating state when the chemical potential is tuned to either of the two Dirac points, shown by the dotted lines in Fig. 1b. The first Dirac point corresponds to a filling of the lowest miniband, or two electrons per site, with corresponding density $n = \frac{4}{\sqrt{3} a^2}$. The second Dirac point corresponds to a filling of the lowest four minibands, or eight electrons per site, with density $n = \frac{16}{\sqrt{3} a^2}$. In both cases only the two bands closest to the chemical potential are relevant to the physics we discuss. We assume the chemical potential is tuned to the second Dirac point, since in this case the bulk gap is larger, and topological features of the system are enhanced.

We may understand the topology of the artificial lattice by constructing a low energy Hamiltonian valid near the Dirac points. For simplicity, we retain only the isotropic terms, setting $H' = 0$ in (\ref{hamil3D}), and consider an infinite square well $V(z)$ of width $d$. Taking a projection onto the lowest confined mode, $ p_z \rightarrow \langle p_z \rangle \rightarrow 0 \ \ , \ \ p_z^2 \rightarrow \langle p_z^2\rangle = \frac{\pi^2}{d^2}$, we find that in the 2D limit the dominant spin-dependent interaction in the Hamiltonian is $\propto - \langle p_z^2 \rangle S_z^2$, and  the spin states in the low-energy sector have $S_z = \pm \frac{3}{2}$. The remaining spin-dependent terms have the form $p_+^2 S_-^2$, $p_-^2 S_+^2$ and induce a momentum-dependent mixing between the low-energy spin states $|\pm \frac{3}{2}\rangle$ and those in the high-energy sector $|\pm \frac{1}{2}\rangle$. Thus we may introduce an effective spin-$\frac{1}{2}$ operator $\bm{s}$ acting on the doublet $|\uparrow\rangle= |\frac{3}{2}\rangle$, $|\downarrow \rangle = |-\frac{3}{2}\rangle$. Near the $K, K'$ points the orbital structure of the Bloch wavefunctions consist of superpositions of basis states $|\sigma \rangle = |a\rangle, |b\rangle$ with
\begin{gather}
|\sigma \rangle = \frac{1}{\sqrt{3}} ( e^{i \tau_z \bm{K}_1 \cdot \bm{r}} + e^{- \frac{ 2\pi i}{3} \sigma} e^{i \tau_z \bm{K}_2 \cdot \bm{r}} + e^{ \frac{2\pi i}{3} \sigma} e^{i \tau_z \bm{K}_2 \cdot \bm{r}}) \ \ ,
\end{gather}
with $\tau_z = +1$ ($-1$) for the $K$ ($K'$) points. Accounting for the magnetic field, the effective Hamiltonian is given by
\begin{gather}
H = -v \tau_z \bm{q} \cdot \bm{\sigma} + 2 \eta \sigma_z s_z - \beta_- \sigma_+ s_+ - \beta_+ \sigma_- s_-
\end{gather}
where the low-energy band parameters are $v = \frac{ 4\pi}{3m a}$,  $m$ is the effective mass of 2D holes, and $2\eta$ is the size of the gap in the topological insulating state, which can generally be either  positive or negative at the second Dirac point and must be calculated from the band structure, but approximately depends on the lattice spacing as $\eta \sim (\frac{d}{a})^4$.  The magnetic field appears via the parameter $\beta_\pm = \frac{ 16 \kappa \mu_B B_\pm}{3} (\frac{d}{a})^2$. Here $\bm{\sigma}$ are the Pauli matrices acting on the orbital doublet $(|a\rangle, |b\rangle)$ (so $\sigma_z |a\rangle = |a\rangle, \ \sigma_z |b\rangle = -|b\rangle$).

\begin{figure*}[t]
	\begin{tabular}{ccc}
		\includegraphics[width = 0.3\textwidth]{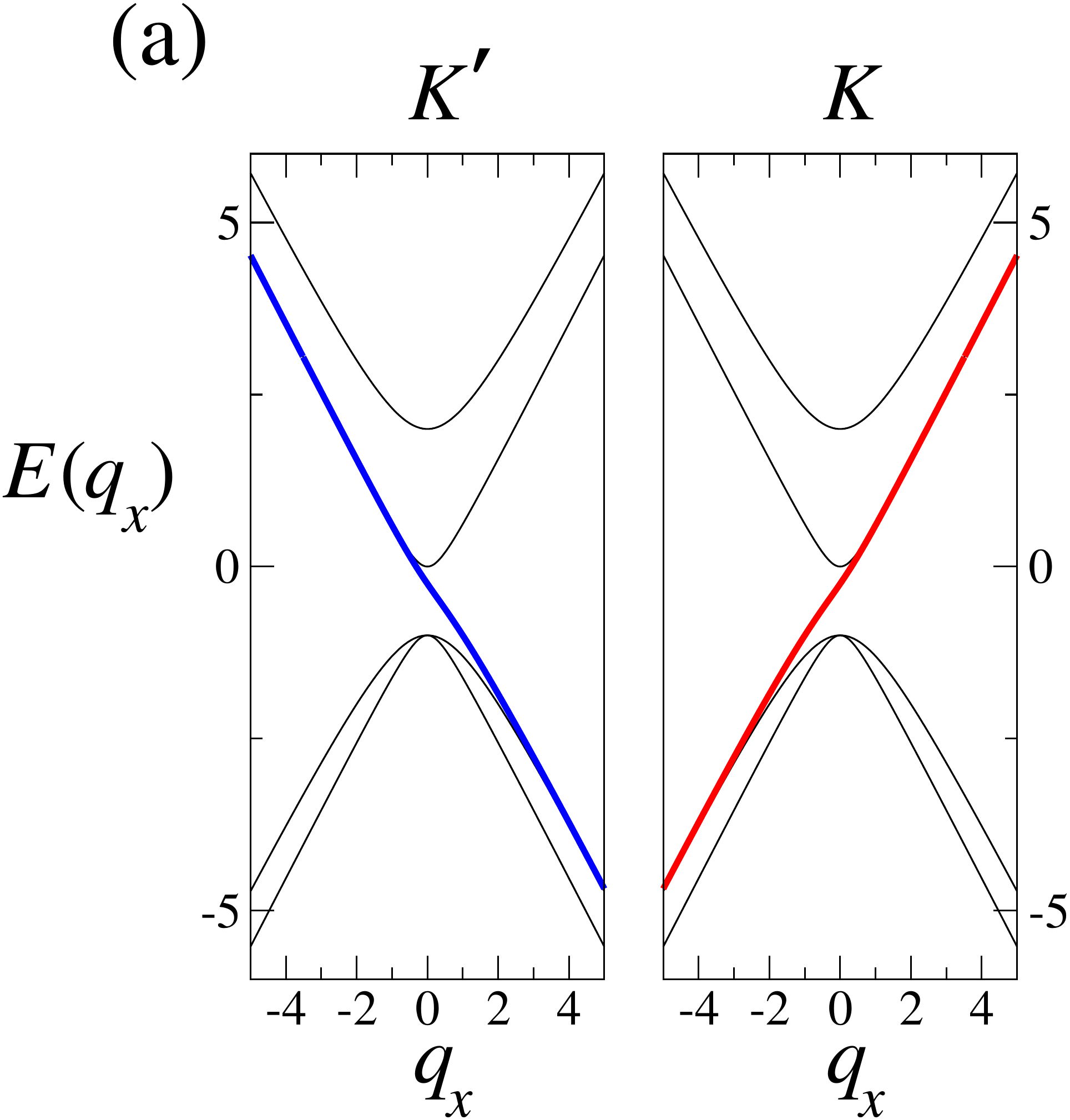} &
		\includegraphics[width = 0.3\textwidth]{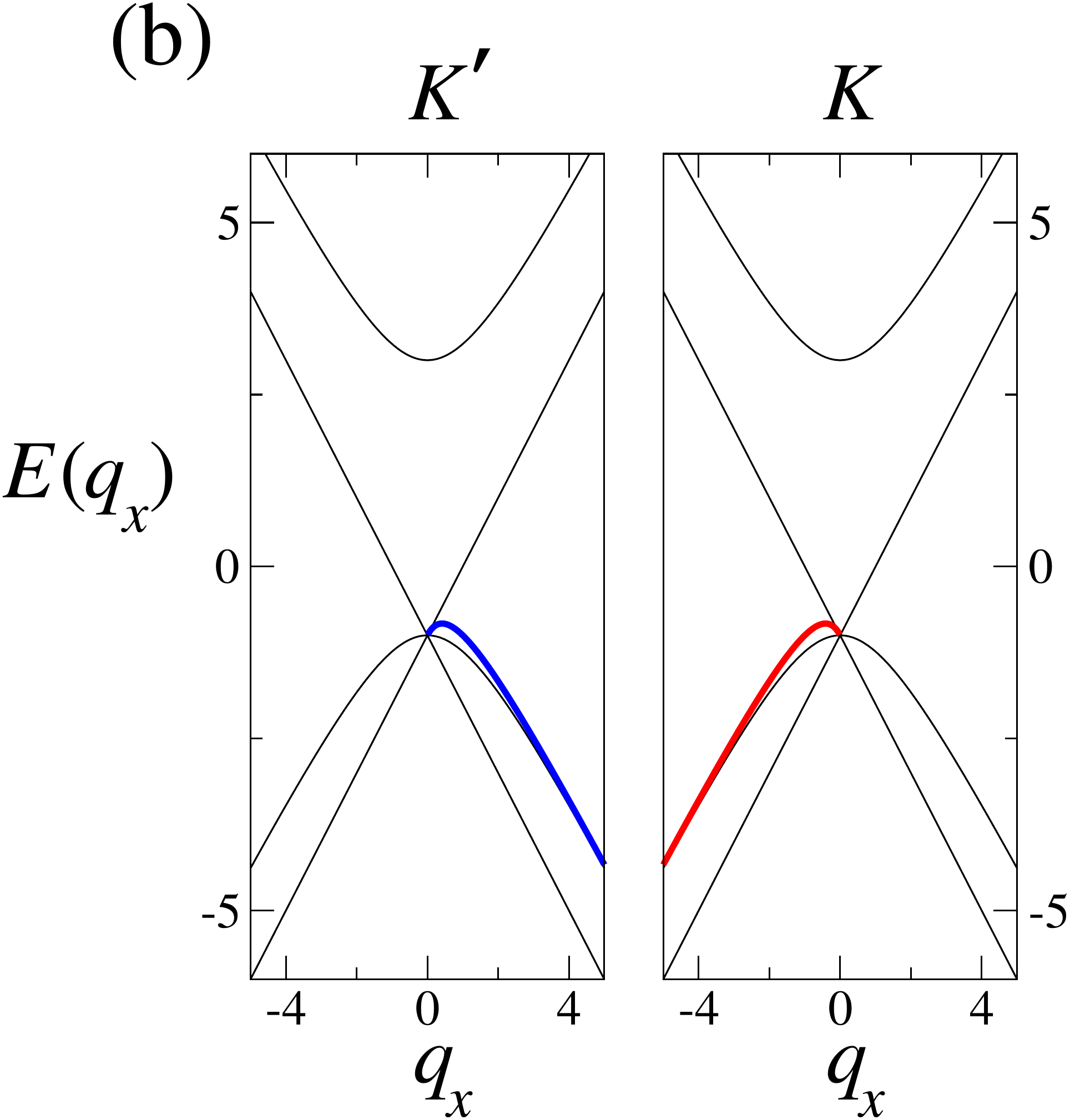} &
		\includegraphics[width = 0.3\textwidth]{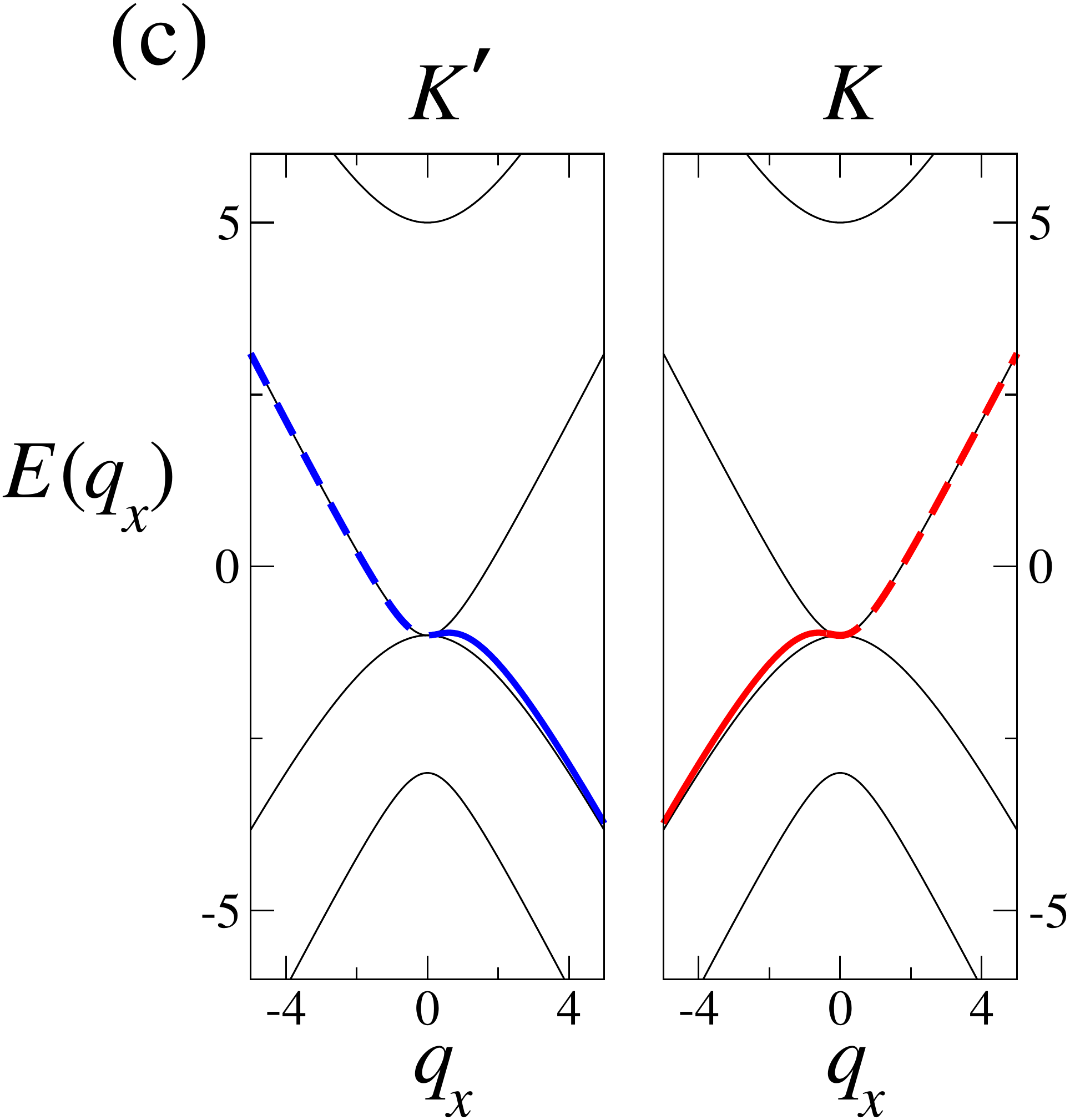}
	\end{tabular}
\caption{ The bulk (thin black lines) and edge (bold red and bold blue lines) dispersions for three distinct values of (a) $\beta = 0.5\eta$, (b): $\beta = \eta$  (the critical state), (c): $\beta = 2\eta$. The dashed  lines in (c) indicate that the branches of the dispersion with $q_x > 0$ at the $K$ valley and $q_x < 0$ at the $K'$ valley are significantly less localized than the branches drawn in solid lines. The units of the axes are chosen so that $\eta = v = 1$.} 
\end{figure*}

Introduction of the magnetic field leads to three qualitatively distinct bulk dispersions illustrated by the black curves in Fig. 2. At small magnetic fields $\beta = \sqrt{ \beta_x^2 + \beta_y^2} < \eta$, the two lowest bands touch quadratically at ${\bm q} = 0$, while the bulk gap $\eta - \beta$ reduces as the magnetic field is increased (Fig. 1a). At a critical magnetic field $\beta = \eta$, the three lowest bands become degenerate at $\bm{q} = 0$ and the bulk gap is closed (Fig. 2b). Above the critical magnetic field $\beta > \eta$, a band inversion occurs and the bulk gap remains closed at the $K$ and $K'$ points where the inner two bands touch (Fig. 2c). Detailed analytical and numerical calculations show that the application of a perpendicular magnetic field, $\delta H \propto - B_z S_z$ reopens the bulk gap at the band-touching points, leading to Chern insulating states which are associated with an integer topological invariant (the Chern number) $N$ which is negative for $B_z > 0$ and positive for $B_z < 0$. In this situation  $P\widetilde{\mathcal{T}}$ symmetry is violated by the perpendicular magnetic field. At $B_z = 0$ the system is in a $P\widetilde{\mathcal{T}}$-invariant semimetallic state which is tuned to the boundary between two topologically distinct insulating phases. Unlike graphene-like 2D Dirac semimetals, which are gapped by an arbitrarily small spin-orbit interaction, the band-touching points in the topological semimetal phase of our hexagonal superlattice cannot be removed by arbitrary spin-orbit interactions at $B_z = 0$, although their location in the Brillouin zone may be shifted. The system remains a topological semimetal in the presence of arbitrary spin-orbit interaction unless the system undergoes a band inversion, e.g. by tuning the magnetic field to cross the  critical point.

It is straightforward to demonstrate the topological robustness of the band touching points in the semimetallic state $\beta > \eta$, by reducing the four-band Hamiltonian to a two-band $\bm{k}\cdot\bm{p}$ model, which may be expressed in terms of the Pauli matrices $L_x, L_y, L_z$ acting on the degenerate doublet of $\bm{q} = 0$ states $(\psi_1, \psi_2)$ at the band touching points. To quadratic order in $\bm{q}$ the Hamiltonian is given by
\begin{gather}
H_{\text{eff}.} = ( \frac{1}{\beta + \eta} - \frac{1}{\beta - \eta} ) \frac{q^2}{4} \nonumber \\
+ \frac{1}{8} ( \frac{1}{\beta + \eta} + \frac{1}{\beta - \eta} )(q_+^2 L_- + q_-^2 L_+) \ \ .
\label{Heff}
\end{gather}
We note that this model is identical to the low energy description of bilayer graphene\cite{BLG}, although crucially in our system the Kramers degeneracy is lacking. The quantum states are given by
\begin{gather}
\psi_{q, \pm} = \frac{1}{\sqrt{2}} ( \psi_1  \pm e^{2i \theta} \psi_2) \ \ ,
\label{wf}
\end{gather}
where $\bm{q} = (|q|\cos \theta, |q| \sin \theta)$ and exhibit an internal rotation of $4\pi$ around the degeneracies. The degeneracies therefore carry a Berry flux of $2\pi$, which is similar to the case in the previously proposed topological semimetal phase of an optical lattice\cite{OptLatt}.

So far, we have only considered the 2D limit of the isotropic terms in the Hamiltonian (\ref{hamil3D}) in a symmetric confinement. The effect of additional terms (and  asymmetry of the confining potential) can be incorporated into the low-energy model via a generic Hamiltonian
\begin{gather}
H_{\text{eff.}} = h(q) + d_x(q) L_x + d_y (q) L_y + d_z(q) L_z
\end{gather}
as long as the system does not contain a triple-degeneracy (e.g. when tuned exactly to the critical point). Under the action of $P\widetilde{\mathcal{T}}$, we have $L_x \rightarrow L_x \ \ , \ \ L_y \rightarrow L_y \ \ , \ \ L_z \rightarrow - L_z$. Thus the $P\widetilde{\mathcal{T}}$ invariance of the full Hamiltonian implies that $d_z(q) = 0$, so that any additional terms can only generate corrections in the Hamiltonian in the operators $L_x, L_y$, and therefore can only shift the location of the band touching points without gapping them. We note that the general case, it is possible to introduce a topological invariant characterizing the band-touching points which does not rely on the quadratic two-band model (\ref{Heff}). This point is discussed in the Supplementary Material.

\begin{figure}[h]
\includegraphics[width = 0.4\textwidth]{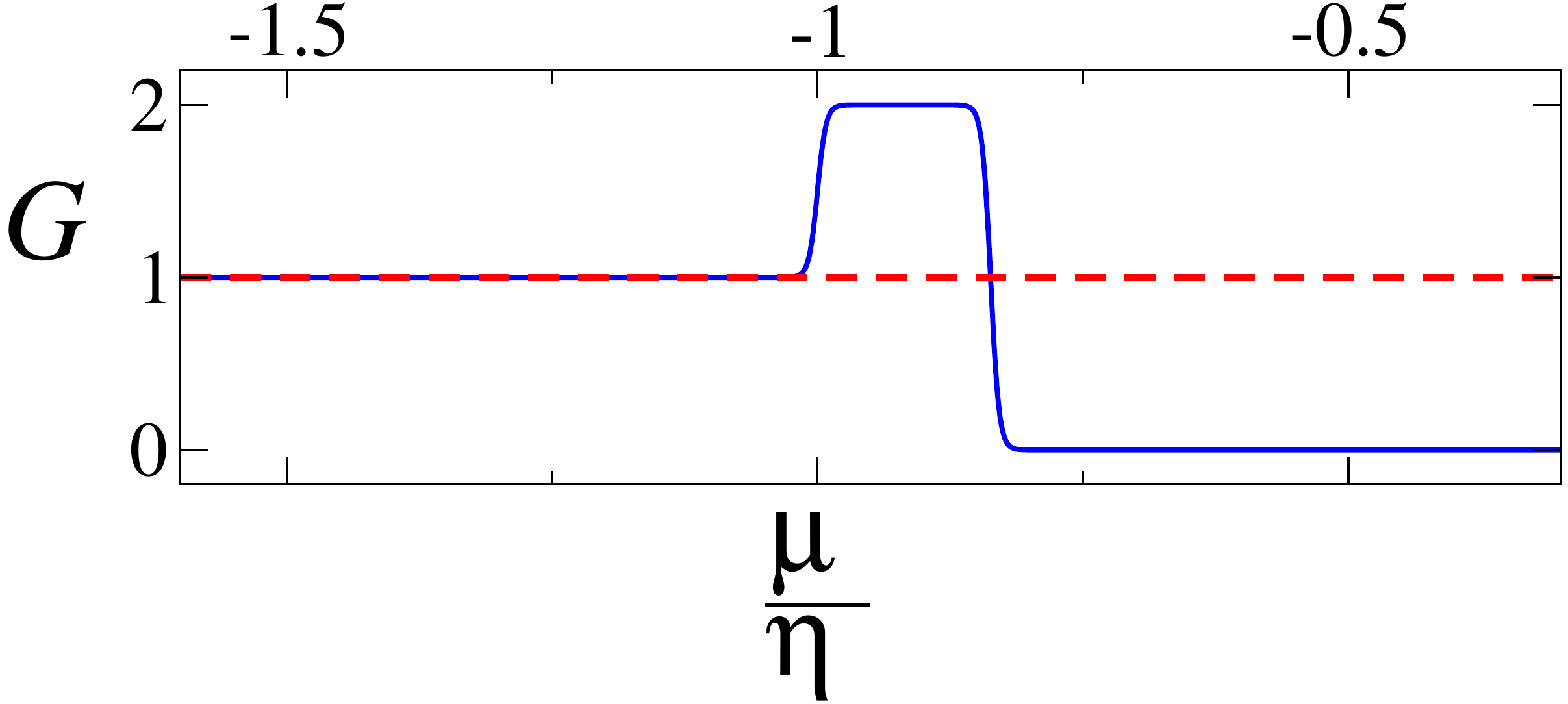}
\caption{The edge conductance in units of $\frac{e^2}{h}$ as a function of chemical potential $\mu$ near the band touching point, $\mu \approx - \eta$ for $\beta \gtrsim \eta$ (blue, solid) $\beta < \eta$ (red, dashed).}
\end{figure}

The quantum phase transition between the insulating and topological semimetallic states can be detected  via the edge modes, which we have calculated analytically for a hard-wall potential normal to the $x$-direction, $U(y > 0) = 0 \ \ , \ \ U(y < 0) = +\infty$. The edge dispersions are plotted in the red and blue curves in Fig. 2 together with the bulk dispersions. The sub-critical states $\beta < \eta$ possess one linearly dispersing mode at each valley (Fig. 2a). As the magnetic field is tuned to the  critical point $\beta = \eta$, the edge states on one side of the $K,K'$ points disappear, with 1D modes existing only for $q_x < 0$ at the $K$ valley and $q_x > 0$ at the $K'$ valley (Fig. 2b). The localization length of the disappearing edge modes tends to infinity as the system approaches the  critical point, so that they gradually become absorbed into the bulk states. The evolution of the edge modes near the  critical point leads to an anomaly in the edge conductance shown in Fig. 3. At magnetic fields away from the critical point, the localization length $L$ of the edge states is given by
\begin{gather}
L \approx L_0=  \frac{v}{\eta} = \frac{3a}{2\pi} \frac{ E_0}{\eta} \ \ , \ \ E_0 = \frac{ 2h^2}{9ma^2} \ \ ,
\end{gather}
thus the edge states are localized and contribute $\frac{e^2}{h}$ to the conductance as long as $\eta$ is not too small compared to the bandwidth $E_0$ (Fig. 3a). When tuned to the  critical point $\beta = \eta$, the edge conductance exhibits steps as a function of the chemical potential $\mu$ (Fig. 3b). The edge dispersions have a maximum $E_{max}$ slightly above the band touching point, and in the energy range $-\eta < \mu < E_{max}$ two edge modes exist at each valley, thus the edge conductance is $G = 2\frac{e^2}{h}$. At energies $\mu < - \eta$, one chiral mode per valley intersects the Fermi energy, giving a conductance of $G = \frac{e^2}{h}$. Above the maximum, $\mu > E_{max}$, the edge modes disappear, and the edge conductance drops sharply to zero.

At supercritical magnetic fields $\beta > \eta$, the branches of the edge dispersion which disappeared at the critical point reappear, but with a large localization length, $\frac{ L}{L_0} \sim 10$ for $\beta = 2\eta$, which suggests that they may be destroyed by hybridization with the bulk states in the presence of disorder. These less localized branches are indicated by the dashed lines in Fig. 1c. While we have plotted only the edge contribution to the conductance, we note that the bulk contribution exhibits a smooth minimum when the chemical potential is tuned close to the band touching point\cite{BLG}, so that the sharp drop in the edge conductance is visible above a smooth background contribution from the bulk conductivity. This striking behaviour of the terminating edge states, which is  reminiscent of the Fermi arc surface states in 3D Weyl semimetals\cite{Weyltheory} recently observed in ARPES measurements\cite{Weylobs}, may be considered the signature of the topological phase transition in our system.

In order to observe the conductance anomaly associated with the delocalization of edge states near the topological phase transition, it is necessary for the localization length $L \sim \frac{E_0}{\eta}$ to be small in the noncritical r\'{e}gime. When $\frac{E_0}{\eta} \sim 10$, the localization length is equal to a few lattice spacings. Considering a GaAs system with  lattice constant $a = 80\text{nm}$, the band parameters are $E_0 = 0.52\text{meV}$ and $\eta = 0.03\text{meV}$ for $d = 20\text{nm}, W = 3E_0$; the topological phase transition occurs at $B_{crit} = 3.6\text{T}$ for a field applied along the $x$-direction. Away from the critical point, the localization length is $L_0 = 4.3a = 0.34\mu\text{m}$, which is an order of magnitude smaller than the mean free path $l \approx 3\mu{m}$ in ultrahigh mobility GaAs hole systems\cite{Chen}. At magnetic field close to the critical point, $B = B_{crit} (1 \pm 0.05)$, the localization length in the disappearing edge modes increases by three orders of magnitude, so that they are fully absorbed into the bulk spectrum. Despite the small value of $\eta$, observation of the topological phase transition remains straightforward via the robust behaviour of the edge conductance.

The ability to create topologically protected semimetallic phases in nanopatterned semiconductors reflects the versatility of these systems afforded by their strong and tunable spin-orbit interactions. Furthermore, it offers exciting possibilities for further studies of spin transport, as well as for the expansion of the topological phase diagram via the introduction of superconducting pairing or strong electron-electron interactions.

\end{document}